\documentclass[aps,pre,preprint,nofootinbib,groupedaddress,showpacs,showkeys]{revtex4}
\usepackage{epsfig}

\begin{document}

\title{Spin glass behavior of the antiferromagnetic Ising model on a scale-free network} 

\author{M. Bartolozzi$^1$}
\email{mbartolo@physics.adelaide.edu.au}
\author{T. Surungan$^{1,2}$}
\email{tasrief@unhas.ac.id}
\author{D. B. Leinweber$^1$}
\email{dleinweber@physics.adelaide.edu.au}
\author{A. G. Williams$^1$}
\email{anthony.williams@adelaide.edu.au}
\affiliation{$^1$Special Research Center for the Subatomic Structure of Matter (CSSM),
 The University of Adelaide, Adelaide, SA 5005, Australia \\ 
$^2$Department of Physics, Hasanuddin University, Makassar 90245, Indonesia}

\date{\today}

\begin{abstract}
Antiferromagnetic Ising spins on the scale-free  Barab$\acute{\rm
a}$si-Albert network are studied via the Monte Carlo method.  Using
the replica exchange algorithm, we calculate the temperature
dependence of various physical quantities of interest including the
overlap and the Binder parameters. We  observe a transition between a
paramagnetic phase and a spin glass phase and estimate the critical
temperature for the phase transition to be $T \sim 4.0(1)$ in units of
$J/k_B$, where $J$ is the coupling strength between spins and $k_B$ is
the Boltzmann constant. Using the scaling  behaviour of the Binder
parameter,  we estimate the scaling exponent to be $\nu \sim 1.10(2)$.
\end{abstract}

\maketitle

\section{Introduction}

In the last few years, the study of complex networks has found
relevance in  various fields including sociology,  ecology, biology,
economics and physics.  In these networks, vertices do not have
homogeneous links or connectivities.  A particularly relevant
structure found in several empirical studies is the so-called {\em
scale-free network} (SFN),  which is characterized by the power law
distribution of the degree of connectivities, $P(k) \sim k^{-\gamma}$,
with $k$ the number of links for a node,  and $\gamma$ the decay
exponent of the distribution.   A network with  $\gamma \rightarrow 0
$  has nodes with a relatively homogeneous number of links (somewhat
resembling the case on regular lattices), while large $\gamma$
corresponds to the existence of ``very famous'' nodes (or hubs), i.e.,
those having direct links to the majority of vertices.

Many networks realized in Nature show scale-free structure. Some
examples studied include food webs~\cite{food_web}, power grids and
neural networks~\cite{watts98,amaral00},  cellular
networks~\cite{cellular}, sexual contacts~\cite{sexual},  Internet
routers~\cite{internet}, the World Wide Web~\cite{www},  actor
collaborations~\cite{watts98,albert99,amaral00,act}, the citation
network of scientists~\cite{citations} and the stock
market~\cite{market}.


 In addition to the scale-free behaviour, these networks are
 characterized by 
a high clustering coefficient, $C$, in comparison
 with random graphs~\cite{bollobas}. The clustering coefficient, $C$,
 is computed as the average of local clustering, $C_i$, for the $i^{\rm th}$
 node, defined as
\begin{equation}
 C_i=\frac{2y_i}{z_i(z_i-1)},
\end{equation} 
where $z_i$ is the total number of nodes linked to the site $i$ and $y_i$  is
the total number of links between those nodes.  As a consequence
both  $C_i$ and $C$ lie in the interval [0,1].
The high level of clustering found
supports the idea that a {\em herding} phenomenon is a common feature
 in social and biological communities.
The parameter $C$ also represents the density of triangles, that is of
elementary cells, associated with the network.

Numerical studies on SFNs have demonstrated how topology
plays a fundamental role in infection spreading~\cite{pastor01}, opinion
formation in large communities~\cite{bartolozzi05}  and
tolerance against random and preferential node removal~\cite{bartolozzi05,tolerance}.
A detailed description of the progress in this emerging field of
statistical mechanics can be found in the recent reviews of 
Refs.~\cite{albert02, dorogovtsev02, dorogovtsev03}.

The aforementioned empirical findings have inspired physicists to investigate
the dynamics of standard models in the new case where the interactions
between elements are described by complex networks.  These include
the study of various magnetic models such as the Ising model.  An
intriguing issue concerns how the unusual topology acts to influence
the cooperative behaviour of the spins.  Studies of the ferromagnetic
(FM) Ising model on a SFN, using several theoretical 
techniques~\cite{Aleksiejuk02,Dorogovtsev02b,Igloi02,Herrero04} including the Monte
Carlo (MC) method~\cite{Herrero04}, have found the robustness of
ferromagnetic ordering against thermal fluctuations for the degree
distribution exponent $\gamma \leq 3$. 
%
%
This result is actually intuitive if we notice that, as $\gamma$
 gets smaller, nodes at the edge of the network will generally have more
 connections. In this situation, the system resembles
the FM Ising model on a regular lattice which exceeds the  
%
%
lower critical spatial dimension, $d_l= 2$.  There the
ordered phase is very robust against thermal fluctuations.  However,
for the antiferromagnetic (AF) case with a SFN, the situation is
different.

Two factors come to play a central role in the dynamics of the AF-SFN
model; namely the competition induced by the AF interaction in the
elementary triangles of the network and the randomness related to the
non-regular connections.  The abundance of elementary triangles in the
network leads to frustration, as, for example, only two of the three
spins can be anti-aligned.  More generally, frustration refers to the
inability of the system to remain in a single lowest energy state
(ground state).  These ingredients lead the AF SFN to belong to a
class of randomly frustrated systems commonly referred to as spin
glasses (SGs).

Most studies of SGs have been performed on regular lattices.  These
studies have shown that frustration and randomness are the key
ingredients for SG behavior, characterized by a frozen random spin
orientation at low temperatures~\cite{SG}.
Spin glasses on a SFN with mixed AF and FM bonds
 have been investigated  recently
by Kim {\it et al.}~\cite{Kim05}. They found, for $\gamma \le 3$ and 
even distributions of the two kinds of interaction, 
that the system is always in a SG state for any finite temperature. 
A study of the pure AF Ising model on a SFN is of great theoretical
interest since, despite the homogeneity of the bonds,
 it inherits all the characteristics of a
SG from the random frustration related to its geometry.
 General reviews on SG systems can be found in Refs.~\cite{SG}.

In this paper we consider the AF Ising model  on a SFN, more precisely
the Barab$\acute{\rm a}$si-Albert (BA) network with tunable
clustering~\cite{Holme02}.  Using
the replica exchange algorithm~\cite{hukushima96} of the Monte Carlo method, 
 we calculate
the order parameters of spin glass behaviour, the so-called
overlap parameter and its distribution.  For an accurate determination
of the critical temperature,  we also evaluate the Binder parameter.
The paper is organized as follows: Section \ref{two} describes the
model and the method. The results are discussed in
Section~\ref{three}. Section~\ref{four} is devoted to  the concluding
remarks.

\section{Model and Simulation Method}\label{two}

\subsection{The model}

\begin{figure}
\centerline{\epsfig{figure=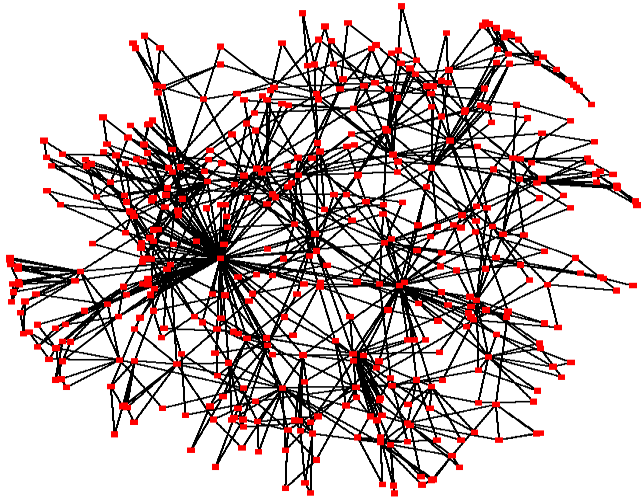,height=7cm, width=9cm}}
\caption{(Color online). Example of a scale-free network. The number of nodes
 is 500 with clustering probability $\theta=0.9$ and $m_0=m=2$.
 The number of nodes has been
kept small in order to preserve the clarity of the plot. Note that, for 
such small networks, a large scale invariant range is obtained only 
if one considers the ensemble average over several realizations.
This plot has been realized with the Pajek software~\cite{pajek}.}
\label{netplot}
\end{figure}

In order to create the scale-free network topology we make use  of the
Barab$\acute{ \rm a}$si-Albert model~\cite{albert99}. This is based on two
main considerations: (i) linear growth and (ii) preferential attachment.
In practice the network is initialized with $m_0$ disconnected
nodes. At each step a new node with $m$ edges is added to the
pre-existing network.  The probability that an edge of the new node is
linked with the $i$th node is expressed by $\Pi(k_i)=k_i/\sum_{j}k_j$.
The iteration of this preferential growing process
yields a scale free network, where the probability of having a node
 with $k$ connections is $P(k)\sim k^{-\gamma}$ with $\gamma= 3$.
%
%
This is an interesting value. In the thermodynamic limit,
 the second moment of the distribution diverges, 
$\langle k^2 \rangle = \infty$, for $\gamma \le 3$.
This leads to peculiar properties of theoretical models in
 this range of $\gamma$ values~\cite{dorogovtsev03}.
In the present work we focus
 on the case in which $\gamma= 3$ and the divergence of
 $\langle k^2 \rangle$ is logarithmic. An extensive
investigation of the phase space for the AF model on SFN is
 left for future work.

It is also worth noting that the Barab$\acute{ \rm a}$si-Albert model  cannot
reproduce a high clustering coefficient. In fact, the value  of this
coefficient depends on the total number of nodes, $N$, in the
network~\cite{albert02} and in the thermodynamic limit, $N \rightarrow
\infty$, $C  \rightarrow 0$.

In the AF Ising system the average cluster coefficient, $C$, plays a
fundamental role in the dynamics. In fact, it represents the average
number of triangles per node and, as a result, it is directly related to
the degree of frustration in  the network. In order to keep this
parameter constant, on average,  with the size of the network, we
introduce a further step in the growth process, namely the triad
formation proposed by Holme and Kim~\cite{Holme02}. In this case, if
the new added node is linked with an older node, $i$, having other
links, then with a certain probability, $\theta$, the next link of the
new node, if any remain, will be added to a randomly selected
neighbour of node $i$.  This method of introducing friends to friends,
while preserving the scale-free nature of the networks with $\gamma
\sim 3$, generates high clustering coefficients that do not depend on
$N$.  The only tunable parameter  that changes the value of the
clustering   coefficient is the  {\em clustering probability}
$\theta$. An example of a SF network generated with this algorithm is
shown in Fig.~\ref{netplot} for 500 nodes. 

We simulate various sizes of the network with many different realizations
and investigate the scaling behaviour of the various physical quantities we are
interested in. All the simulations have been carried out fixing
$\theta=0.9$, corresponding to an average clustering coefficient of $C
\sim 0.39$, close to the value found in many real systems
\cite{albert02}.
On each SFN constructed at the beginning of the simulation, we assign
to each vertex an Ising spin, and to each link an AF interaction.  The
Hamiltonian can be written as follows
\begin{equation}\label{ham}
  H = -\sum_{\langle ij \rangle} J_{ij}\,  s_i\, s_j \, .
\end{equation}
Here the summation is performed over the connected spins $s_i$ and
$s_j$ occupying sites $i$ and $j$, respectively. The coupling
interaction $J_{ij} = J =-1$ is AF.
As previously mentioned, each vertex with the local  cluster
coefficient $C_i > 0$ together with its neighbours, compose elementary
triangles.  Due to the AF interactions the local system is frustrated. 

It is worth pointing out that $C$ is related to the degree of
frustration of each network.  Due to the probabilistic algorithm used
for their construction, the value of $C$ fluctuates around a mean value 
from one network to the next and, therefore, provides a source
 of randomness that, as we will
see, gives rise to the spin glass properties of the model.
  This probabilistic growth is not shared by other algorithms which use
recursion formulas to generate scale-free structures, such as, for example,
 the Apollonian networks~\cite{Andrade05}. In this case, once one fixes the number
of iterations of the algorithm, which is proportional to the number of nodes 
of the final network, one also fixes its topology. 
The element of randomness is therefore missing in the Apollonian procedure.

As a random system, each realization of a network of size $N$ will
differ in the ``structure'' of connectivities.  Therefore, in order to
have reliable statistics, we average over many realizations of the SF
network for each specified size.  The system sizes that we
simulate are $N =$ 1024, 2048, 4096, and 8192.  
In general, one takes into account
more realizations for small system sizes and less for large system
sizes as the latter tend to self-average. However,  since the
self-averaging of physical quantities for larger system sizes is
interfered with by the increase of ground state degeneracy, we do not take
less realizations. Instead all physical quantities of interest for
each system size are averaged over 1000 network realizations.
%
%
Moreover, for each realization of the network, we fix $m_0=m=5$, corresponding
to a coordination number on a regular lattice of approximately 10. 
In the thermodynamic limit, the average connectivity for 
the BA network is $ \langle k \rangle = 2 m = 10$,
emphasizing the fact that we are implicitly
dealing with a high dimensional system. 

Another peculiarity of SF networks is the existence of a broad
distribution of ``hubs'', that is nodes with a large number of
connections, $k$.  The energy difference in a spin flip actually
depends on the number of connections of the spin itself, $\Delta
E_{i}= -2 s_i \sum_{j=1}^{k_i}s_j$.  Thus in the AF case for the $i$th
spin with $k_i$ connections, the hubs are more likely to ``freeze''
into a particular configuration compared to the nodes with just a few
links.  This property resembles the spin glass behaviour of particular
alloys where some elements freeze into a particular orientation at a
higher temperature than others.

\subsection{Simulation method}

The calculation of the thermal averages of the physical
quantities of interest is performed using the replica exchange 
MC method~\cite{hukushima96}.  In this method the
evolution of $M$ replicas, each in equilibrium with a heat bath of
inverse temperature  $\beta_m$ for the $m^{\rm th}$ replica, is simulated in
parallel.  Given  a set of inverse temperatures, $\{ \beta \}$, the
probability distribution of finding the whole system in a state $\{ X \} =
\{ X_1,X_2, \dots, X_M\}$ is
\begin{equation}
 P(\{ X, \beta\}) = \prod_{m=1}^{M} \tilde{P}(X_{m},\beta_{m}),
\end{equation}
with
\begin{equation}
\tilde{P}( X_m, \beta_m) = Z(\beta_{m})^{-1} \exp(-\beta_{m} H(X_{m})),
\label{equil}
\end{equation}
and $Z(\beta_m)$ is the partition function at the $m^{\rm th}$ temperature.
We can then define an exchange matrix between the replicas in our Markov chain,
 $W(X_m,\beta_m| X_n,\beta_n)$, that is the probability
to switch the configuration $X_m$ at the temperature $\beta_m$
with the configuration $X_n$ at $\beta_n$. By using the detailed balance condition, 
required to keep the entire system at equilibrium, on the transition matrix
\begin{eqnarray}
P( \ldots,\{ X_m, \beta_m \},\ldots, \{ X_n, \beta_n \},\ldots )\cdot
 W(X_m,\beta_m| X_n,\beta_n) \nonumber \\ = P( \ldots,\{ X_n, \beta_m \},
\ldots, \{ X_m, \beta_n \},\ldots )
\cdot W( X_n,\beta_m | X_m,\beta_n), 
\end{eqnarray}
along with Eq.~(\ref{equil}), we have that 
\begin{equation}
\frac{ W( X_m,\beta_m | X_n,\beta_n)}{ W( X_n,\beta_m | X_m,\beta_n)}=\exp(-\Delta),
\end{equation}
where $\Delta=(\beta_{n}-\beta_{m})(H(X_{m})-H(X_{n}))$.
With the above constrains we can choose the matrix coefficients 
according to the standard Metropolis method and, therefore, we have
\begin{equation}
W(X_m,\beta_m| X_{n},\beta_n)=\left \{  \begin{array}{ccc} 
1 & {\rm if} & \Delta<0, \\ 
\exp(-\Delta) & {\rm if} & \Delta>0.
\end{array} \right.
\label{trans}
\end{equation}
In our simulation we restrict the exchange to temperatures next to
each other; that is, we consider only the terms $W(X_m,\beta_m|
X_{m+1},\beta_{m+1})$.  This choice is motivated by the fact that  the
acceptance ratio decays exponentially with $(\beta_n-\beta_m)$.

The replica exchange method is extremely efficient for simulating
systems such as spin glasses,  that can otherwise become frozen in
some particular configuration at low temperatures when using a
standard Metropolis algorithm for the configuration update.  In this
case, as we lower the temperature, the system can become trapped
into a local minimum of the free-energy where the barriers are so high
that the time required for the system to move to another allowed
region of the configuration space diverges to infinity as a function
of the system size.  If the system is trapped in a local minimum then
the ergodicity condition is not fulfilled anymore  and the measure
that one makes become biased by the particular region of the
configuration space that is being sampled.     By using the exchange
replica method, instead, we keep switching the temperatures between
the $M$ copies of the system and, as long as the  higher temperature
is in a hot phase (where, the system can easily explore all the
configuration space), then we are in principle able to explore all the
configuration space also for the lower temperatures. Another advantage
of this method is that the replica exchange reduces drastically  the
temporal correlation in the system dynamics at each temperature. This
enables one to collect more independent measures for the thermal
averages of the physical quantities and, therefore, reduces the
uncertainty.

It is important to stress that, before starting the actual
simulations,  some care is required in selecting the set of inverse
temperatures, $\{ \beta \}$. In fact, the method is efficient only
when a fairly large transition probability is maintained in the range
of interest. From Eq.~\ref{trans}, we can see that, in  the hot phase,
temperatures can be more coarsely spaced while  in the cold phase the
temperatures need to be closer to each other.  An optimal set of
temperatures can be obtained by iterating, in preliminary runs, the
following map~\cite{hukushima96}:
\begin{equation}
\begin{array}{c} 
 \tilde{\beta}_{1}=\beta_1, \\ 
\tilde{\beta}_{m}=\tilde{\beta}_{m-1}+(\beta_{m}-\beta_{m-1})\cdot p_{m}/c,
\label{map}
\end{array}
\end{equation}
where $p_{m}$ is the acceptance ratio for the switch between two
 configurations at the $m$th temperature and
 $c=\sum_{m=1}^{M}p_m/(M-1)$ is a normalization factor. The initial
 value for the set $\{ \beta \}$ is uniform in the interval of
 interest and  we ensure that $\beta_1$ belongs to the hot phase.  For
 each iteration of the map, a run of a few thousand MC steps  is carried
 out to calculate the acceptance ratios, $p_m$, which are then plugged into
Eq.~(\ref{map}) in order to obtain a new set of inverse temperatures.
 After a few
 iterations,  the map of Eq.~(\ref{map}) converges to a fixed point,
 $\{ \beta^{\star} \}$,  which sets the values of the temperatures to
 be used in our simulations. 


 In using this method, we define a ``local'' MC (LMC) update as a MC
update for each spin of each replica, either consecutively through all
elements of the network or randomly.  Given that we can group the
inverse temperatures in even and odd pairs, $(\beta_{m},\beta_{m+1})$,
after each LMC update we alternate attempts to switch configurations
from one temperature to the next.  According to this procedure, we
define a Monte Carlo step (MCS) as a LMC plus a half ($m$ odd or even)
exchange trial.

For each realization of the network we start from a random
configuration of the spins and then  perform $10^3$ LMC updates in
order to reach  thermal equilibrium. After this transient period, we
run the simulation for $3 \times 10^5$ MCSs while taking  a total of
$6 \times 10^4$ measures for the thermal averages, that is one every 5
MCSs (temporal correlations are lost very quickly by using the replica exchange method).
   We consider low temperatures in a search for the possible
existence of a phase transition.  The thermal averages obtained for
each network are then averaged over the ensemble of networks.  In the
following, we indicate $\langle...\rangle$ as the thermal average and
$\left[...\right]_{\rm av}$ as the ensemble average.  The statistical
errors in the plots, where reported, are calculated via the bootstrap
method.

\section{Results and Discussion}
\label{three}

\subsection{Spatial correlations and specific heat}

\begin{figure}
\centerline{\epsfig{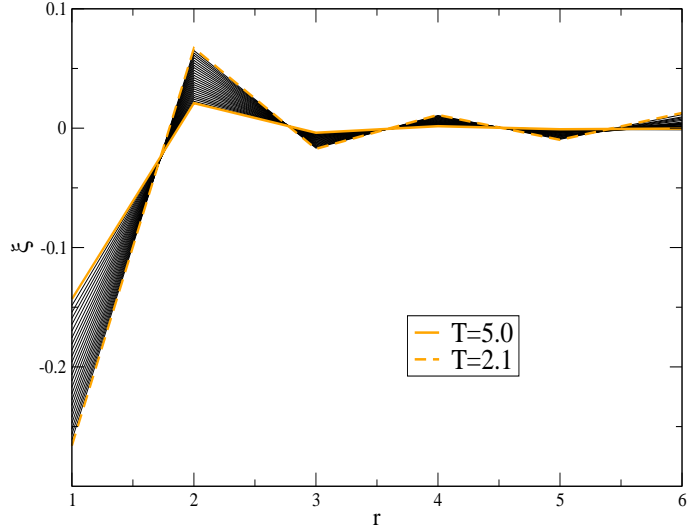}}
\caption{ (Color online). 
Spatial autocorrelation, $\xi(r)$, for $N=1024$ averaged over 50
network configurations for temperatures between $T=5.0$ and $T=2.1$.
 The plot shows that
 next neighbour spins tend to be anti-parallel as in standard AF Ising model. 
The AF interaction in the triangular units of the system results in high frustration.
Note that the number of nodes at large distances is much smaller than the ones 
at smaller distances and so the average calculated for $r=5$ and $r=6$ includes just
few samples. This is a consequence of the ``small-world'' effects in SF networks.} 
\label{sp_corr}
\end{figure}

As a first step we investigate the extent of spatial correlation of the spins in
the SF network by making use of the spatial autocorrelation function which is
defined on a regular lattice as
\begin{equation}
 \xi(r)= \left[ \frac{1}{L_d} \langle  s_i s_{i+r} \rangle \right]_{\rm av},
\label{spat_corr}
\end{equation}
where $L_d$ is the total number of pairs at distance $r$ and 
depends just on the dimension considered. 
In a SF network the situation is more complicated since there may be
several paths leading from a certain node to another.
 We then define $r$ as the {\em minimum}  
path between two nodes and the denominator of the Eq.~(\ref{spat_corr}) 
becomes dependent on $r$. The results, averaged over 50 configurations,
 between the temperatures of $T=5.0$ and $T=2.1$ are 
shown in Fig.~\ref{sp_corr} for $N=1024$. 
All the temperatures in the present paper are expressed in units of  $J/k_{B}$,
 where $J$ is the coupling strength between spins and $k_B$ is the Boltzmann constant.

\vspace{1.5cm}
\begin{figure}
\centerline{\epsfig{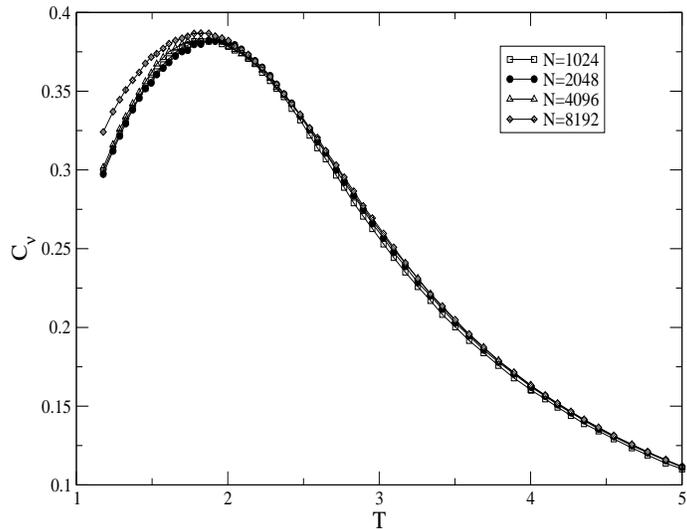}}
\caption{Specific heat, $C_{\nu}$, as a function of the temperature and system size.
The plot has been obtained by averaging over
50 network configurations for each $N$. Note that the specific heat does not scale
with the size of the system.}
\label{sp_heat}
\end{figure}

In order to give a better interpretation of the plot in Fig.~\ref{sp_corr} we
remind the reader about an important
propriety of SF networks; that is their ``small world structure''.  The
``hubs'', in fact, play a fundamental role in linking  sites
otherwise very distant. Moreover, the average path length increases
just logarithmically with the size of the network~\cite{albert02,dorogovtsev02}.  In
the plot of Fig.~\ref{sp_corr}, for $N=1024$ nodes, an upper limit of $r=6$ is encountered. While all the 50 configurations reach $r=6$, only a few networks exceed this limit.  

The plot emphasizes how neighboring  spins, on average, tend to be anti-correlated, as expected
in the AF case. The autocorrelation decreases with the
distance from the node under consideration.  The temperature
dependence is also in accord with the expectations. The absolute value of the correlation
decreases  with increasing  temperature and vice versa. Indeed, 
the highest and lowest temperatures form a perfect boundary for all
the curves. 
This is an expected result, since thermal effects always tend to
 reduce the correlation between the spin interactions.

We also study the behaviour of the specific heat, $C_{\nu}$, defined as
follows
\begin{equation}
 C_{\nu}(T)= \left[ \frac{1}{Nk_{B}T^2}(\langle E^2 \rangle - \langle E \rangle^2) \right]_{\rm av},
\end{equation}
where $k_{B}$ is the Boltzmann constant.  Although no singularity is
expected for this quantity in the spin-glass transition, 
it is interesting to compare its behaviour with other studies.
The dependence of the specific heat on temperature 
is reported in Fig.~\ref{sp_heat}. The statistical errors, in this case,
 are smaller than the size of the symbols and therefore are not reported. 
 A common Schottky peak of the specific heat for a finite system is observed  at the temperature of $T \sim 2.0$ independent 
of the system size.
Below this point, we found that $C_\nu$  decreases and goes to
zero as $T \rightarrow 0$.
 
This behaviour follows from simple entropy considerations. 
In fact, since we are dealing with a finite Ising system,
the entropy is bounded at each finite temperature as well,
\begin{equation}
 S(T)=\int_{0}^{T}\frac{C_{\nu}(T)}{T}dT < 2^{N},
\end{equation}
and, necessarily, $C_{\nu} \rightarrow 0$ for $T \rightarrow 0$.   
 
The next section is dedicated to study of the SG
behaviour and the phase transition of the system. In order to achieve
this task, we evaluate the corresponding order parameters, the overlap
parameter and the Binder parameter.

\subsection{Observing spin glass behaviour}

With the presence of frustration and randomness in the AF-SFN model,
we expect to observe a spin glass transition, i.e., a transition from
a temporal disordered to a temporal ordered phase at low temperatures. 

This feature is not shared by the so-called fully frustrated
systems~\cite{tasrief}.
This type of transition might  be characterized by the  order
parameter such as that suggested by Edward and Anderson~\cite{EA},
defined as follows 
\begin{equation}
q_{EA} = \left[ \frac{1}{N}\sum_i\langle s_i \rangle^{2} \right]_{\rm av}.
\end{equation}
However, an ergodic Markov chain of a system having $Z_2$ symmetry will ensure the thermal average of the $i$th spin vanishes. Therefore a finite value of this measure simply reflects the non-ergodicity in the MC update.  


A more appropriate quantity that is often used to characterize the SG state is the
overlap parameter, $q$,  defined as~\cite{parisi,bhattY}
\begin{equation}\label{qorder}
q =  \frac{1}{N}\sum_{i} s_i^{(\alpha)} s_i^{(\beta)},
\end{equation}
where the superscripts $\alpha$ and $\beta$ denote two copies of the
same configuration of connectivity at the same temperature. 
The actual value of $q$ is extracted from both the thermal 
and disorder average, $ \left[ \langle... \rangle \right]_{\rm av}$.

Using the replica exchange MC simulation, the two copies,  
 $\alpha$ and $\beta$, are allocated at each temperature of 
the parallel tempering.
 This means, if the measurement is performed on $M$ points of
temperatures, there are $M$ pairs of replicas.
 The Metropolis spin update is performed on each node for every MC step.
 As a part of the equilibration steps of the algorithm described in the previous 
section, we exchange two 
 $\alpha$ (and $\beta$) replicas of neighboring temperatures, 
according to a certain probability. 
Then, for each temperature, the $\alpha$ and $\beta$ replicas  are 
superimposed every 5 MCSs in order to measure the overlap parameters, 
as defined in Eq.(\ref{qorder}).

In particular, for the Ising system, due to the $Z_2$ symmetry, it is
important to evaluate the absolute value of the order parameter,
\begin{equation}
|q| \equiv \left[ \langle |\frac{1}{N}\sum_{i} s_i^{(\alpha)} s_i^{(\beta)}| 
\rangle \right]_{\rm av},
\end{equation}  
to overcome the implication of the $Z_2$ symmetry of the Hamiltonian,
that is the configurations ${s_i}$ and ${-s_i}$ have equal Boltzmann
weights.  That is, if the system is at thermal equilibrium and if we
take quite long MCS then the usual $q$ should average to zero. The
existence of a spin glass phase is indicated by the convergence of
$|q|$ to a finite value as we increase the network size. At the same
time, a convergence of $|q|$ to zero at high temperatures is
anticipated.  In the latter case the system is in the paramagnetic
phase.

\begin{figure}
\centerline{\epsfig{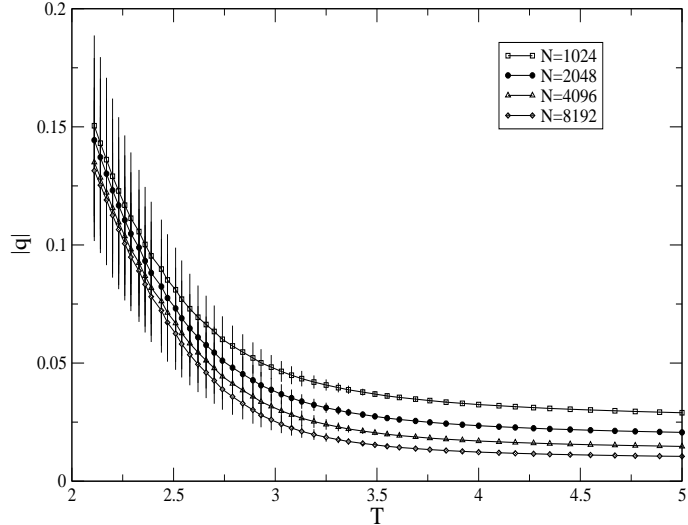}}
\caption{Temperature dependence of the overlap parameter, $q$,
for different system sizes $N$.  The increasing value of $q$ at low
temperatures indicates a SG phase.  For a given network size, 1000
realizations of the SFN are averaged over. }
\label{overlap}
\end{figure}

The temperature dependence of $|q|$, resulting from the simulations,
is shown in Fig.~\ref{overlap}.  The existence of a SG phase is
indicated by the finite value of $|q|$ in the low temperature region,
and the approach of $|q|$ to zero at higher temperatures associated
with the paramagnetic phase.  For high temperatures and large networks,
$|q|$ is approaching zero in accord with the thermodynamic limit where
$|q| = 0$~\cite{Ogielski85}.

\begin{figure}
\vspace{1cm}
\centerline{\epsfig{figure=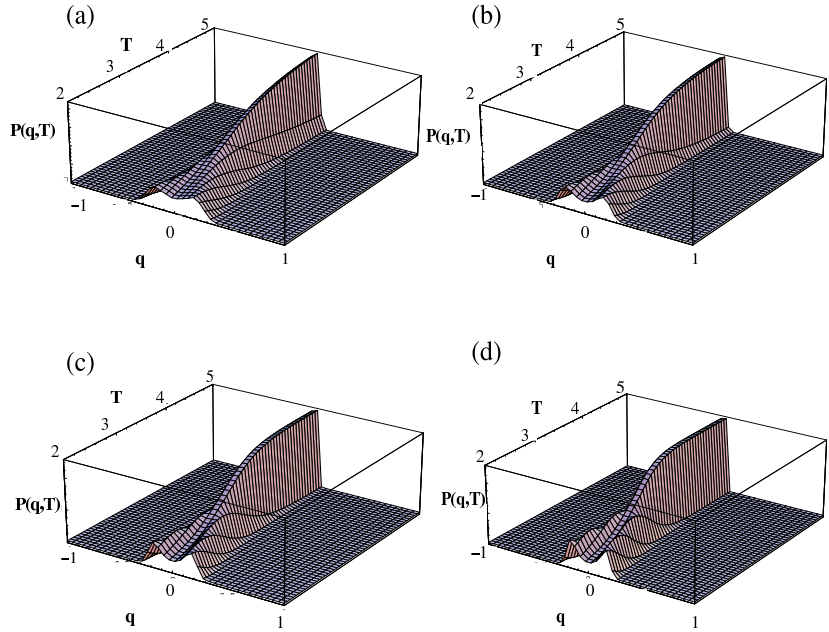,height=9cm, width=11cm}}
\caption{ (Color online). 
The distribution of $q$ at various temperatures for different system sizes,
including (a) $N=1024$, (b) $N=2048$, (c) $N=4096$ and (d) $N=8192$.}
\label{distrib}
\end{figure}

The existence of these two different phases can also be observed from
the distribution of $q$, as shown in Fig.~\ref{distrib}.  For
higher temperatures we observe simple Brownian fluctuations of the
values of $q$, leading to a singly peaked Gaussian distribution
characteristic of a paramagnetic state.  By decreasing the
temperature, the distribution spreads out, reflecting the
increasing number of metastable disordered states associated with 
a substantial frustration.  At lower temperatures the
distribution develops double peaks reflecting the $Z_2$ symmetry and a
 finite value of $|q|$, representative of the SG phase.  We note that the
shape of the observed distribution at low temperatures is different from that of the
conventional Ising system where the double peaks approach delta-like
double peaks reflecting a simple doubly degenerate ground state~\cite{dotsenko}.

An accurate evaluation of critical temperature of the phase transition
is achieved via the Binder parameter defined as follows
\begin{equation}
g_L = \frac{1}{2}\left(3 - \frac{\left[\langle q^4\rangle\right]_{\rm
av}}{\left[\langle q^2\rangle \right]_{\rm av}^{2}}\right).
\end{equation}  
Here $\langle q^2 \rangle$ and $\langle q^4 \rangle$ are respectively
the second and the fourth cumulant moment of $q$. In this calculation,
 in order to avoid
systematic correlation errors that could bias the results if we were
evaluating this average  over $g_L$ directly~\cite{Kawashima96}, the
second and fourth order cumulants are averaged prior to taking their
ratio. The Binder parameter is constrained in the range  $0 \le g_L
\le 1$.
At high temperature, where thermal fluctuations overcome all
cooperative interaction, the system is expected to exist in the
paramagnetic phase where there is no spatial  autocorrelation.  As a
result, the distribution of $q$ should be Gaussian centered at $q=0$.
In this case the ratio of the cumulants, $\langle q^4 \rangle /\langle
q^2 \rangle^2 \rightarrow 3 $, resulting in $g_L \rightarrow 0$.
At low temperatures, the cooperative interaction becomes dominant and
the ratio of the cumulants approaches unity so that $g_L \rightarrow 1$.

\begin{figure}
\centerline{\epsfig{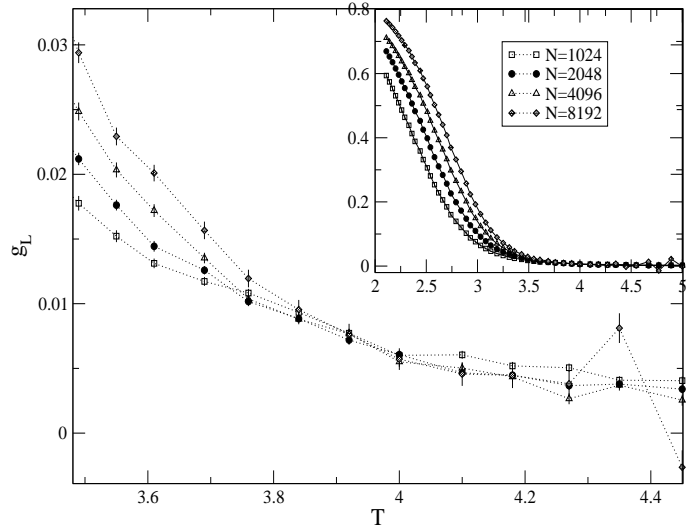}}
\caption{Scaling behaviour of the Binder cumulant, $g_L$, for
  different system sizes.  Each system size is averaged over 1000
  realizations of the network configuration.}
\label{binder}
\end{figure}

Fig.~\ref{binder} (inset) displays the temperature dependence of the Binder
parameter for a variety of network sizes.  A spin-glass state is
observed for lower temperatures where the Binder parameter deviates
from zero, and increases with the system size while approaching to 1. 
 In the thermodynamic limit, we expect $g_L \to 1$ just below the critical
temperature.  A crossing point in the size dependence of $g_L$
indicates that the critical temperature for the SG phase transition is
$T \sim 4.0$.
Fig.~\ref{binder}  indicates that for temperatures above $T \sim 4.0$
the Binder parameter, while remaining always above zero, does indeed
order in an opposite manner indicative of a genuine crossing of the
curves and in accord with a genuine spin glass transition at finite
%
%
temperature. This  feature which is not observed for 
uniformly distributed AF and FM bonds, as $T_c = \infty$
 in the thermodynamic limit~\cite{Kim05}. 
However, the value of the transition temperature is not  
determined with high accuracy by the crossing of the Binder parameter. 
In fact, finite size effects seem to slightly distort the tendency 
for very small networks, as in the case of $N=1024$. At the same time,
the statistical errors in the paramagnetic phase for large networks, see $N=8192$,
 appear to be significant and some points are scattered. 

\vspace{1cm}

\begin{figure}
\vspace{1cm}
\centerline{\epsfig{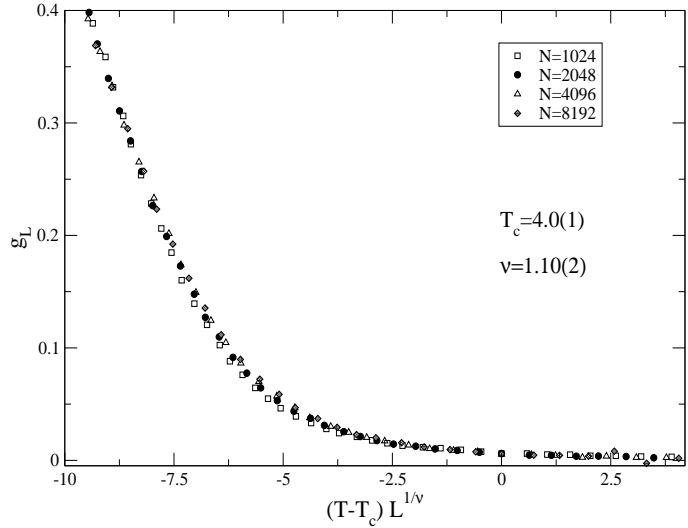}}
\caption{Scaling plot of the data illustrated in Fig.~\ref{binder},
 fitted to Eq.~\ref{scalebind}.}
\label{fig_bind}
\end{figure}

A more accurate estimate of the critical temperature, $T_c$, for
finite size systems can be obtained using scaling arguments.  For a SG
system, the Binder parameter depends on the system size
$L$ as

\begin{equation}\label{scalebind}  
g_{L} = \tilde{g}_{L}[(T-T_c)L^{1/\nu}],
\end{equation}
being $\nu > 0$ the spin glass correlation length exponent, implying
that  at $T_c$ the Binder cumulant does not depend on $L$.  For the
SFN, the system size scales logarithmically with the number of nodes
$N$~\cite{albert02,dorogovtsev02,dorogovtsev03,Kim05} and therefore we take
$L = \log(N)$.
This slow increase in the diameter of the system, as well as the
 average path length, is a manifestation of the ``small-world''
 property of this network, induced by the presence of a large number
 of highly connected hubs which  create shortcuts between the nodes.
 An important implication of this feature is that we cannot embed the
 network in any finite dimensional lattice: we are implicitly dealing
 with a high dimensional system.
The correlation length, in this case, is still well defined although its value
gets close to the densely-connected, mean field limit as we increase the
 average connectivity of the nodes, $\langle k \rangle =2 m$.  


The parameters $T_c$ and $\nu$ are determined by constraining the
temperature dependence of the Binder parameter for each network size
to lie on a single curve.  The curves following the scaling bahaviour
of Eq.~(\ref{scalebind}) are shown in Fig.~\ref{fig_bind}.
From this fit we estimate the critical temperature $T_c\sim 4.0(1)$
and the exponent of the SG correlation length $\nu \sim 1.10(2)$.
It is important to underline that this kind of behaviour is not observed
for an AF system on a regular triangular lattice. 

\section{Concluding Remarks}\label{four}

In summary, we have investigated the antiferromagnetic  Ising model on
 a Barab$\acute{\rm a}$si-Albert scale-free network  using the replica
 exchange Monte Carlo method.  Through the calculation of  the overlap
 parameter  we observe spin glass behaviour at low temperatures.
 Using the scaling behaviour of the Binder parameter  the critical
 temperature separating the SG and the  the paramagnetic phases is
 found to be $T_c=4.0(2)$  with a scaling exponent of SG correlation
 length $\nu \sim 1.10(2)$. Such behaviour is not observed for the AF
 Ising model on regular triangular lattices. Hence the topology of the
 interactions plays a critical role in the dynamics of the system.

\section*{Acknowledgments}

The authors wish to thank Y. Okabe, E. Marinari and J.-S. Wang 
for valuable discussions.  
One of the authors (TS) is grateful for  the hospitality of the Center for the
Subatomic Structure of Matter (CSSM) at the University of Adelaide
during his academic visit to the Center.  The computation of this
work has been done using the Hydra teraflop supercomputer 
facility of the South
Australian Partnership for Advanced Computing (SAPAC).

\end{document}